\documentclass[]{JHEP}
\def\Im{\mathop{\mbox{Im}}}
\def\Re{\mathop{\mbox{Re}}}
\def\Tr{\mathop{\mbox{Tr}}\,}

\def\di{\mbox{d}}
\newcommand{\bea}{\begin{eqnarray}}
\newcommand{\be}{\begin{equation}}
\newcommand{\eea}{\end{eqnarray}}
\newcommand{\ee}{\end{equation}}
\newcommand{\nn}{\nonumber}

\title{D-particles in the space-time of the shock wave}

\author{Ciprian Acatrinei, Marco Fabbrichesi and Roberto Iengo\\
INFN, Sezione di Trieste and\\
Scuola Internazionale Superiore di Studi Avanzati (SISSA)\\
via Beirut 4, I-34013 Trieste, Italy.}

\abstract{We study the interaction of two D-particles in the space-time of
the shock wave. 
We first write the amplitude in string theory 
and find that, at large distances from the shock-wave source, the  $O(v^4)$
term in the relative velocity $v$ is an $\alpha'$-independent function
of the interbrane separation $b$. The amplitude is therefore that of 
supergravity---for large $b$, only closed-string massless modes 
contribute.
We then show how the same result is obtained
in the matrix model  (at small $b$) by setting up the formulation of 
the dimensionally reduced super Yang-Mills
theory in the curved background of the shock wave.}


\preprint{SISSA 1/99/EP \\
January 1999}
\begin{document}
\section{Motivations and Summary}                   

D-particles (D0-branes) are the simplest example of Dirichlet 
branes~\cite{bigbook}. Their 
dynamics has been studied in order to understand their role in
string theory and to compare their behavior with that of gravitons
having momentum in a compactified direction in the dimensionally reduced
$d=11$ supergravity theory (that is, the long-distance limit of M theory).
This comparison is important in identifying which dynamical
configurations are susceptible of a description in terms of light
open-string degrees of freedom~\cite{2,4}, thus providing the effective ``low energy''
theory of D-branes that is at the basis of the matrix model~\cite{BFSS}. 
This issue
has been analyzed both in the case of two-body interactions~\cite{BBPT} and 
in the more arduous case of three bodies~\cite{us,3B} (see also~\cite{DR} for
a related temporary controversy). 

A better understanding of the physics of D-brane in non-flat space-time is
also desirable because  it is a subject relevant to
many open questions in the interplay between string theory and
gravitation. At the moment, very little is known about the matrix model 
in curved spaces~\cite{lit1}.
Recent analyses discuss the case of the Schwarzschild metric~\cite{lit2}.
The present paper is  a first enquire in this direction.
 
One of the  simplest examples of dynamics in curved space is the  motion of 
two D-particles in the so-called shock-wave metric, that is,
the gravitational field generated by a massless particle. 
Since there are two particles in addition to the massless source of gravity,
this is also a case of a three-body interaction, which
is of special interest because of the inherent peculiarities
of the three-body dynamics. Moreover, the shock-wave metric itself
is particularly
important in the study of quantum gravity, see~\cite{HH,AK,HS,FR,FIR}
and, recently, \cite{HI}.

In order to pursue our program, we approach the problem 
from two different points of view.

First, we perform an analysis in terms of
closed-string theory.
Since the simplest massless particle in this theory is the graviton,
we take it to be the source for the shock-wave metric. 

We write the amplitude
for the scattering of two D-particles and two gravitons describing the
incoming and outgoing massless source.  The process in which
we are interested emerges as
a {\it pinching} limit of the full amplitude, where the two graviton 
vertex operators collide in the world-sheet, implementing within
the string algorithm  
the emission of the long-range gravitational field (the shock-wave) from the
incoming-outgoing source.
The full amplitude in this pinching limit describes  the interaction of
two D-branes with themselves and with the shock-wave.

In this computational
framework, the D-particles are special boundary states at the ends of a
cylindrical world-sheet, and the string amplitude is obtained by computing
the correlator of the two vertices on the cylinder (we consider the
ten dimensional uncompactified space-time). The D-particles are treated
in the usual eikonal approximation, however the computation of the
amplitude is not quite conventional because of the
peculiarities of the three-body kinematics.
After summing over the spin-structures we look at the expansion in the
 relative velocity $v$ of the D-particles. 

The first result of our work is that, up to the fourth order in 
$v$, the leading singularities in the momentum transfer from the
source-gravitons are a function of the interbrane distance $b$ that is 
the same for arbitrary $b$ and therefore 
 independent from $\alpha'$, see eq.~(\ref{last}). 
This result generalizes well-known properties of brane
interaction to the case of a non-trivial background.
The leading singularities in the momentum transfer correspond to the
leading powers of the transverse distance $r_\top$ from the 
shock-wave source.
In particular, the fourth order term in $v$ gets contribution
from the leading behavior of the shock-wave gravitational field,
that is $O({r_\top}^{-6})$,
whereas the second order term is sub-leading and $O({r_\top}^{-8})$.
We will focus on the fourth order term.
 
Of course, the long interbrane-distance regime
is dominated by the massless modes of the closed string and
is therefore the supergravity (and M theory) result.

On the other hand, we can consider the same process in the
short interbrane-distance regime where we expect  the theory 
to be described by the dimensionally reduced
$d=10$ super Yang-Mills theory corresponding to the exchange of massless
open-string states.
The gauge group  is $SU(2)$ because we want to describe the
interaction of two D-particles. The novelty is here that the 
super Yang-Mills
 theory must be
written in a gravitational background. 
This frames the problem in the language of the
matrix model, for which the D-particles are the fundamental degrees
of freedom, in the rather unconventional case of a curved background. 
The final amplitude 
describes the motion of two D-particles in the space-time of the 
shock wave. 

We show how, by performing the small $v$ and large ${r_\top}$ expansion
of the matrix-model result,
the amplitude of the closed-string 
computation is reproduced, see eq.~(\ref{ultima}).

\section{String theory} 

Consider two D-particles located at 
$\vec Y_{1}^{\bot}$ and $\vec Y_{2}^{ \bot}$ 
and moving with velocities $\vec v_1$
and $\vec v_2$, respectively, where 
$\vec Y_{(i)}^{\bot} \cdot \vec v_{(i)}=0$. We consider the frame where 
$\vec v_1+\vec v_2= \vec Y_{1}^{\bot}+\vec Y_{2}^{\bot}=0$. We call 
$\vec v=\vec v_1-\vec v_2$ and 
$\vec b=\vec Y_{1}^{ \bot}-\vec Y_{2}^{ \bot}$, where by definition
$\vec b \cdot \vec v=0$.
Their interaction with an external gravitation field---generated by 
the source graviton
moving along the direction $z$ at a transverse distance $\vec r_\top$ 
(by definition $\vec r_\top$ is orthogonal to  $z$)
from the center of mass of the two D-particles---is dominated by a term
proportional to $1/q^{2}$, where $q$ is the momentum
transfer by the gravitation field 
to the system of the two D-particles.
The momentum $\vec q$ can be separated
into a part  $q_z$ along the direction of motion of the
external graviton (i.e., the shock-wave direction), 
and the remaining orthogonal directions, 
$\vec q_\top$. In the eikonal approximation, $q_0 = q_z$ that implies
$q^2 = q^2_\top$.

In the frame chosen, the motion of the center of mass of the
particles is factorized out. The
effect of the external source graviton on it is taken into account
by disconnected diagrams---which we do not consider---in which 
the source couples independently
first to one of the two D-particles and then to the other. Here 
we focus instead
on the relative motion of the two D-particles as they interact
with the gravitational field.

We write the amplitude in $d=10$ as
\be
a\, (r_T,b,v) 
= \int \frac{\di q_z \, \di ^{d\mbox{-}2}\vec q_\top}
{(2\pi)^{d-1}} \:e^{i \vec r \cdot \vec q_\top} 
\int \di \ell \:\int \di^2 w\; \di^2 z\;\, {\cal A}\, (z,w;\,\ell) \, ,
 \label{a}
\ee 
where (up to an overall constant)
\be
{\cal A}\, (z,w;\,\ell) = \sum_S\:(\pm)_S\: 
\langle v_2, Y_2 | e^{- \ell H}\:
  V_k (z, \bar{z})\, V_p (w, \bar{w})   | v_1, Y_1 \rangle_{S} \label{A}
\ee
contains the interaction of the two D-particles with
the two gravitons, the vertices of which are given by
\be
V_p (z, \bar{z}) = \varepsilon_{\mu\nu}^{(p)} \Bigl[ \partial X^\mu(z) + 
i p \cdot \psi (z) \psi^\mu(z) \Bigr] \Bigl[ \bar{\partial} X^\nu(\bar{z}) + 
i p \cdot \bar{\psi} (\bar{z}) \bar{\psi}^\nu(\bar{z}) \Bigr] 
e^{ip\cdot X} \, ,
\ee
and similarly for the other one at $w$ and carrying momentum $k$. 
The index $S$ stands for the three even
spin structures of the fermionic propagators. The amplitude (\ref{A}), in the
case of $\vec v=0$, has
been considered in~\cite{pasquinucci}.

For later use, we define the two vectors $q$ and $P$ to be
\be
q_0 = k_0+p_0\, , \quad  \vec q = \vec k + \vec p  \quad \mbox{and} \quad
 P_0 = k_0-p_0 \, , \quad   \vec P = \vec k - \vec p \, .
\ee

The D-particle kets or bras are given by a tensor product of
center-of-mass and string-oscillator states.
The D-particle center-of-mass state in space-time  
is described by a Fourier transform in terms of the 
eigenstates of the energy-momentum transfer $Q$. 

One has further to integrate over 
the world-line of the D-particle, which in the eikonal
approximation is the straight line $Y^{\mu}(t)$ parameterized by 
$Y(t) = (t, \vec R(t))$ with   
$ \vec R = \vec v\, t  + \vec Y_{\bot}$. We thus get 
\be
 | v, Y_\bot \rangle = \int \di t
\int \frac{\di^{d}  Q}{(2\pi )^d} \,
e^{i \,(Q_0t-\vec Q\cdot \vec R(t))} | v, Q \rangle \, 
=\int \frac{\di^{d\mbox{-}1} \vec Q}{(2\pi )^{d\mbox{-}1}} \, 
e^{i \,\vec Q\cdot \vec Y_\bot} | v, Q_0=\vec v\cdot\vec Q,\vec Q
 \rangle \, ,
\ee
Notice that in the eikonal approximation we have the constraint
 $Q_{0i} = \vec v_i \cdot \vec Q_i \equiv v_i Q_L^i$. 
This is similar to what holds for the momentum transfer from the 
source graviton (whose velocity is equal to 1 along the $z$-direction) 
namely $q_0 = q_z$.

In eq.~(\ref{A}) the 
Hamiltonian $H$ transports the D-particle at $\vec Y_2^\bot$ and is given by
\be
H = H_{\rm \scriptscriptstyle CM} + H_{\rm \scriptscriptstyle OS} = 
\frac{1}{2} \, Q_2^2 + H_{\rm \scriptscriptstyle OS} \, ,
\ee
where $H_{\rm \scriptscriptstyle OS}$ contains the string oscillators.

Let us also define the normalized correlators
\be
\langle  V_k (z, \bar{z})\, V_p (w, \bar{w}) \rangle_{S} = 
\frac{ 
\langle v_2, Y_2 | e^{- \ell H}\:
  V_k (z, \bar{z})\, V_p (w, \bar{w})   | v_1, Y_1 \rangle_{S}}
{\langle v_2, Y_2 | e^{- \ell 
H_{\rm \scriptscriptstyle OS}} | v_1, Y_1 \rangle_{S}} \, ,\label{corr}
\ee
where
\be
\langle v_2, Y_2 | e^{- \ell H_{\rm \scriptscriptstyle OS}} | v_1, Y_1 \rangle_{S} =
\frac{\vartheta_S \left( iv/\pi\right)\vartheta_S^3 \left( 0\right)}
{\vartheta_1^{'4} \left( 0\right)} \, , 
\ee
and therefore
\be
{\cal A}\, (z,w;\,\ell) = \sum_S\:(\pm)_S\: 
\langle  V_k (z, \bar{z})\, V_p (w, \bar{w}) \rangle_{S}
\: \frac{\vartheta_S \left( iv/\pi\right)\vartheta_S^3 \left( 0\right)}
{\vartheta_1^{'4} \left( 0\right)} \, .
\ee

Since we are not interested in spin effects,
we can consider only the term proportional to 
$\varepsilon^{(p)} \cdot \varepsilon^{(k)}$. The two-graviton correlation
function is then
\bea
\langle  V_k (z, \bar{z})\, V_p (w, \bar{w}) \rangle_{S} & = &
\varepsilon^{(p)}_{\mu\lambda}\varepsilon^{(k)}_{\nu\rho} \: \Bigl[ 
\langle \partial X^{\mu} \partial X^{\lambda} \rangle 
\langle \bar{\partial} X^{\nu} \bar{\partial} X^{\rho} \rangle 
\Bigr.   \label{vv} \\
&- &  \langle \psi^{\mu} \psi^{\lambda} \rangle_S 
\langle \bar{\psi}^{\nu} \bar{\psi}^{\rho} \rangle_S
\langle p \cdot \psi \, p \cdot \bar{\psi}\,  k \cdot \psi \, k \cdot 
\bar{\psi}
\rangle_S \nn \\
& +&  \langle \partial X^{\mu} \partial X^{\lambda} \rangle 
\langle \bar{\psi}^{\nu} \bar{\psi}^{\rho} \rangle_S 
 \langle p \cdot \bar{\psi} \, k \cdot \bar{\psi} \rangle_S \nn \\
&+ &   \Bigl.
\langle \bar{\partial} X^{\mu} \bar{\partial} X^{\lambda} \rangle
 \langle \psi^{\nu} \psi^{\rho} \rangle_S
\langle p \cdot \psi \, k \cdot \psi \rangle_S \Bigr] 
 \langle e^{i k\cdot X(z)}\, e^{i p\cdot X(w)} \rangle \, . \nn
\eea

We look for the singularities as we take the limit
$z \rightarrow w$:
\bea
\langle  V_k (z, \bar{z})\, V_p (w, \bar{w}) \rangle_{S} & \rightarrow&
\frac{\varepsilon^{(p)} \cdot \varepsilon^{(k)}}{(4 \pi)^2}  \left\{
\frac{1}{|z-w|^4} \left( 1 + O(q^2) \right)   \right. \nn\\
&&   - \frac{1}{|z-w|^2}   
\Biggl[ \langle k \cdot \psi \, k \cdot \bar{\psi} \rangle_S  
 \langle p \cdot \psi \, p \cdot \bar{\psi} \rangle_S \label{vv2}\\
& &  + \left. \Biggl.
\langle k \cdot \psi \, p \cdot \bar{\psi} \rangle_S  
 \langle k \cdot \bar{\psi} \, p \cdot \psi \rangle_S  
 + O(q^4) \Biggr] \right\}  
\langle e^{i k\cdot X(z)}\, e^{i p\cdot X(w)} \rangle \, . \nn
\eea
As we shall see, we can neglect in (\ref{vv2}) $O(q^2)$ terms in the
quartic pole and  $O(q^4)$ terms in the quadratic pole. 

 While
the second term, which comes from the fermionic correlators, is already a
quadratic pole, the first one, which comes from the bosonic correlators,
is quartic and, in order to contribute, it must be multiplied by 
terms containing a factor $|z-w|^2$.
In fact, it is
the quadratic pole times a factor $|z-w|^{q^2/(4\pi)}$, contained in
$\langle e^{ik\cdot X} e^{ip\cdot X} \rangle$, 
that gives rise to the singularity $1/q^2$ for which we 
are looking:
\be
\int \di^2 (z-w)|z-w|^{-2 +q^2/(4\pi)} = \frac{8\pi^2}{q^2} \, . 
\ee

We split the center-of-mass modes $X_0$ from the string-oscillator modes
in the correlator:  
\be
\langle e^{i k\cdot X(z)}\, e^{i p\cdot X(w)} \rangle =
\langle e^{i k\cdot X_0(z)}\, e^{i p\cdot X_0(w)} \rangle 
\langle e^{i k\cdot X(z)}\, e^{i p\cdot X(w)} 
\rangle_{\rm \scriptscriptstyle OS} \, ,
\ee
and analyze their contributions separately.

\subsection{String center-of-mass modes}

Let us first consider the string center-of-mass
modes. They are given by
\be
X^\mu_0(z) = X^\mu_0 - i Q^\mu \Im z \, ,
\ee
where $\Im z$ plays the role of proper time of closed string
propagating from one D-particle to the other one.

Because
\bea
\langle \vec Q_2, Q_2^0 | e^{i k\cdot X(0)} =
\langle \vec Q_2 -\vec k, Q_2^0 -k_0 |  \nn \\
 e^{i p\cdot X(0)} | \vec Q_1, Q_1^0 \rangle =  
|\vec Q_1 +\vec p, Q_1^0 +p_0 \rangle \, ,
\eea
we have two conservation laws given by the sandwiching of the external
states:
\bea
\langle \vec Q_2 -\vec k, Q_2^0 -k_0 | \vec Q_1 +\vec p, Q_1^0 +p_0 \rangle
& = &
\delta^{(d-1)} \left( \vec Q_2 -\vec k - \vec Q_1 - \vec p  \right) \nn \\  
&& \times \: \delta \left(  Q_2^0 - k_0 -  Q_1^0 - p_0 \right) \, .
\eea

Notice that the energy conservation gives 
$\vec Q_2 \cdot \vec v_2 -\vec Q_1 \cdot \vec v_1 = q_z$. 

According to (\ref{corr}), for
\be
Z_{\rm \scriptscriptstyle CM} \equiv
\langle e^{i k\cdot X_0(z)}\, e^{i p\cdot X_0(w)} \rangle =
\langle v_2, Y_2 |e^{-\ell  H_{\rm \scriptscriptstyle CM}}
\: e^{i k\cdot X_0(z)} e^{ip \cdot X_0(w)}  | v_1, Y_1 \rangle \, ,
\ee
we have
\bea
Z_{\rm \scriptscriptstyle CM} &=&
\int   \frac{\di^{d\mbox{-}1} \vec Q_1}{(2\pi )^{D\mbox{-}1}}
\int    \frac{\di^{d\mbox{-}1} \vec Q_2}{(2\pi )^{D\mbox{-}1}}
\exp \left[ i \vec Q_1 \cdot Y^1_\bot -  i \vec Q_2 \cdot Y^2_\bot
\right] \nn \\
&& \times 
\exp \left\{ -\frac{\ell}{2} \left[ \vec Q_2^2 - \left(\vec v_2 \cdot Q_2
\right)^2 \right] + \Im z \left( \vec k \cdot \vec Q_2 - 
k_0 \, \vec v_2 \cdot \vec Q_2  \right) \right. \nn \\
&& \Biggl.   + \Im w \left( \vec p \cdot \vec Q_1 - 
p_0 \, \vec v_1 \cdot \vec Q_1 \right) \Biggr\}
 \delta^{(d)} \left( Q_2 - Q_1 - k - p \right) \, .  
\eea  

We replace by means of the $\delta$-functions
$\vec Q_2 = \vec Q_1 + \vec q$; we set, by neglecting $\vec q \cdot \vec v/v$,
$Q_L = - q_z/v$ and $\vec Q_1 = \vec Q$.
Finally, we integrate  over the components $\vec Q_\bot$
orthogonal to $\vec v$ and find
\bea
Z_{\rm \scriptscriptstyle CM} &=&
-\frac{1}{v} e^{-i \vec q \cdot \vec b/2}\:
\int \frac{\di^{d\mbox{-}2} \vec Q_\bot}{ (2\pi)^{d\mbox{-}2}} \\
&& \times \exp \left[ -\frac{\ell}{2}\, \vec Q_\bot^2 + 
\left( \Im z \, \vec k +
\Im w \, \vec p + i \vec b \right) \cdot \vec Q_\bot \right] \nn \\
&& \times \int \frac{\di q_z}{2\pi}  \exp \left[
   -\frac{\ell}{2} \left[ 1 +  (\Im z  + \Im w)\,\frac{v_z}{\ell} \right]
  \left( \frac{q_z}{v} \right)^2 \right] \, , \nn
\eea
which, after integration in $\vec Q_{\bot}$ and $q_z$, for
small $v_z$ and $q_\top \rightarrow 0$, gives
\be
Z_{\rm \scriptscriptstyle CM} = \ell^{-(d-1)/2} \: 
 e^{\left(  \mbox{\scriptsize Im}\: z \vec k + 
 \mbox{\scriptsize Im}\: w \vec p + i \vec b \right)^2/2 \ell}
\left[ 1 - \frac{v_z}{2 \ell} \left(\Im z  + \Im w
\right) \right]  \label{z3} \, ,
\ee
where we have dropped overall factors of $2\pi$.

 We retain a term proportional
to $|z-w|^2$ from the expansion of the exponential in
(\ref{z3}), and finally obtain
\bea
Z_{\rm \scriptscriptstyle CM}& =&
- \ell^{-(d-1)/2}\:  e^{- \vec b^2/2 \ell} \:
\left\{ 1 - \frac{v_z}{2 \ell} \left(\Im z  + \Im w
\right) \right. \nn \\
& &  + |z-w|^2\left.  \left[ \frac{k_0^2}{4\ell}
-\frac{1}{16 \ell^2} (\vec P \cdot \vec b )^2 
\right] \right\} \, . \label{f41}
\eea

\subsection{Oscillator modes}

The oscillator part is given by the expectation value of the
exponential factors
\bea
\langle e^{i k\cdot X(z)}\, e^{i p\cdot X(w)} 
\rangle_{\rm \scriptscriptstyle OS} & = &
e^{- \langle \left[ k\cdot X(z) + p\cdot X(w) \right])^2 
\rangle_{\rm \scriptscriptstyle OS} /2} \nn \\
&= & \left| \frac{\vartheta_1(z-w)}{\vartheta_1(z - \bar{w})}\right|^{q^2/{4\pi}}
\:  \left| \frac{\vartheta_1^2(z-\bar{w})}
{\vartheta_1(z - \bar{z})\,\vartheta_1(w - \bar{w})}\right|
^{k_{\scriptscriptstyle 0}^2/{4\pi}} \, ,
\eea
which in the $z \rightarrow w$ limit gives
\be
\langle e^{i k\cdot X(z)}\, e^{i p\cdot X(w)} 
\rangle_{\rm \scriptscriptstyle OS} 
\longrightarrow \left[ 1 + |z-w|^2 
 \frac{k_0^2}{2\pi} \, \partial^2_w 
\ln \vartheta_1(w-\bar{w})  \right] 
 \left| 
\frac{\vartheta_1(z-w)}{\vartheta_1(z - \bar{w})}\right|^{q^2/{4\pi}} 
\, , \label{f42}
\ee
where we have kept terms up to $|z-w|^2$ as required.

The fermionic correlator can be written (up to $O(q^2)$) as
\bea
\langle k \cdot \psi \, k \cdot \bar{\psi} \rangle_S 
 \langle p \cdot \psi \, p \cdot \bar{\psi} \rangle_S
& +& \langle k \cdot \psi \, p \cdot \bar{\psi} \rangle_S  
 \langle k \cdot \bar{\psi} \, p \cdot \psi \rangle_S
 =  \nn \\ 
& & \left( \frac{1}{4\pi}\right)^2 \left\{
k_0^2 \frac{(\vec q_\top 
\cdot \vec  v)^2}{v^2} \left[ Q_S^2 - R_S^2 - P_S^2 \right]
+ q^2 k_0^2 \left[ Q_SP_S + P_S^2 \right] \right. \nn  \\
&- & q^2 \:\frac{({\vec P} \cdot {\vec v})^2}{4 v^2}
\left[ P_S^2 - Q_SP_S \right] +
\left. k_0\, q^2\: \frac{{\vec P} \cdot{\vec v}}{v} R_S P_S \right\}
\, , \label{ferm} 
\eea
where the spin-structure dependent functions
$Q_S$, $R_S$ and $P_S$ are given in the appendix.

We perform now the integration
\be
\int \di^2 z \int \di^2 w = 
\int \di^2 (z-w) \int \di \Re w \int \di \Im w \, .
\ee
Every term in (\ref{ferm}) is $O(q^2)$. Unless we find an
additional singularity coming from the integration over $\Im w$, the 
$q^2$ in front of all terms will cancel the $1/q^2$ coming from the
integration over $(z-w)$ and we would be left without the $1/q^2$ pole in
which  we
are interested. We therefore look for
terms behaving like $(w-\bar w)^{-1-q^2/4\pi}$.
After summing over the spin structures, a factor of this form could arise
from terms in (\ref{ferm}) proportional to $Q_S^2$, $R_S^2$, $Q_SP_S$ and
$R_SP_S$. The function $P_S^2$ does not contain the required
term. The contributions coming from the terms proportional
to $Q_S^2$ and $R_S^2$ are equal and therefore cancel.
Therefore, there are only two terms in (\ref{ferm}) that can give rise
to the required power of $v$ and $1/\vec q_\top^{\:2}$:
\bea
\langle k \cdot \psi \, k \cdot \bar{\psi} \rangle_S 
 \langle p \cdot \psi \, p \cdot \bar{\psi} \rangle_S 
& +& 
\langle k \cdot \psi \, p \cdot \bar{\psi} \rangle_S  
 \langle k \cdot \bar{\psi} \, p \cdot \psi \rangle_S \label{f21} \\
 \longrightarrow &&  \left( \frac{1}{4\pi}\right)^2  \left\{  
k_0\, q^2 \:\frac{{\vec P} \cdot{\vec v}}{v} R_S P_S +
 q^2 \left[ k_0^2 + \frac{(\vec P \cdot \vec v)^2}{4 v^2}
\right]  Q_SP_S  \right\} \, . \nn
\eea
The  sum over the spin structures yields, for $v\rightarrow 0$, 
\bea
\sum_S (\pm)_S \,R_S\,P_S\,  
\frac{\vartheta_S\left(-iv/\pi\right)\vartheta_S^3 
\left(0\right)}{\vartheta_1^{'4}(0)}
& = & \frac{1}{2} \sum_S  (\pm)_S\, 
\frac{\vartheta_S\left(w-\bar w \right)\vartheta_S^2 
\left(0\right)}{\vartheta_1^2(w - \bar w)\vartheta_1^{'2}(0)} \nn \\ 
&& \times \left[ e^{v}\: \vartheta_S \left(w-\bar w -iv/\pi\right)
- e^{-v}\: \vartheta_S \left(w-\bar w +iv/\pi\right) \right] \nn \\
&& \longrightarrow \quad \frac{i}{\pi} \frac{v^3}{(2\pi)^2} 
\frac{\vartheta_1'(w - \bar w)}{\vartheta_1(w - \bar w)} \label{RP} \, ,
\eea   
and
\bea
\sum_S (\pm)_S \,Q_S\,P_S\,  
\frac{\vartheta_S\left(-iv/\pi\right)\vartheta_S^3 
\left(0\right)}{\vartheta_1^{'4}(0)}
& = & \frac{1}{2} \sum_S  (\pm)_S\, 
\frac{\vartheta_S\left(w-\bar w \right)\vartheta_S^2 
\left(0\right)}{\vartheta_1^2(w - \bar w)\vartheta_1^{'2}(0)} \nn \\
 && \times \left[ e^{v}\: \vartheta_S \left(w-\bar w -iv/\pi\right)
+ e^{-v}\: \vartheta_S \left(w-\bar w +iv/\pi\right) \right] \nn \\
&& \longrightarrow \quad 2 \left( \frac{i v}{2\pi} \right)^4 
\frac{\vartheta_1''(w - \bar w)}{\vartheta_1(w - \bar w)} \label{RP2} \, , 
\eea
after dropping a total derivative, which does not contribute because the
contributions at the integration limits cancel against each other.

\subsection{The final amplitude}

We can now collect all the relevant terms by putting in evidence
the integral over $z$ around the quadratic pole and
writing everything else for $z=w$. The integrand does not
depend on $\Re w$ and its integration gives a factor 1.
 We thus obtain
\bea
a\, (r_\top ,b,v)& \simeq& - \int \frac{\di ^{d\mbox{-}2}\vec q_\top}
{(2\pi)^{d \mbox{-}1}} e^{i \vec r \cdot \vec q_\top} \:
 e^{- \vec b^2/2 \ell} \:
\int \di \ell \: \ell^{-(d-1)/2} \label{f2f4} \\
&& \times \int \di^2 (z-w) \, |z-w|^{-2 +q^2/(4\pi)}
\int_0^\ell \di \Im w\: \Bigl[ F_2(w,\bar w)+ F_4(w,\bar w) \Bigr] \, .
 \nn
\eea

In eq.~(\ref{f2f4}), $F_4(w,\bar w)$ comes from the bosonic
correlator, the term proportional to $|z-w|^{-4}$ in (\ref{vv}), 
times the terms proportional to $|z-w|^2$ in (\ref{f41}) and 
(\ref{f42}) which compensate for the quartic
pole, and is given by 
\be
 F_4(w,\bar w) = 
\left(  \frac{iv}{2\pi} \right)^4 \left[
\frac{k_0^2}{2\pi}\, \partial^2_w 
\ln \vartheta_1(w-\bar{w})  +  \frac{k_0^2}{4 \ell} - \frac{1}{16} \, 
\frac{\vec b\cdot \vec P}{\ell^2} \right] \, ,
\ee 
where the factor in front comes from the expansion in $v$ of
the $\vartheta_1 (i v/2\pi)$ obtained after summing over the
spin structures (see formulas in the appendix).

The term $F_2(w,\bar w)$ in  eq.~(\ref{f2f4}) originates from the
fermionic correlators, the term proportional to
 $|z-w|^{-2}$ in (\ref{vv}), 
and it has two terms, one that 
originates from  (\ref{RP2}) and one from the product
of the fermionic correlator (\ref{RP}) and the term
proportional to $v_z$ in (\ref{z3}):
\bea
F_2(w,\bar w) &= &
-  2\, \frac{k_0^2 q^2}{(4 \pi)^2} \left[
\left( 1 + \frac{v_z^2}{v^2} \right) 
\left(\frac{iv}{2\pi}\right)^4  
\frac{\vartheta_1''(w - \bar w)}{\vartheta_1(w - \bar w)} \right.\label{q2} \\
& & \left. - \frac{v_z}{v} \left( \frac{i v^3}{4 \pi^3} \right)
\, \frac{\vartheta_1'(w - \bar w)}{\vartheta_1(w - \bar w)}  
\, \left(\frac{v_z}{\ell} \right) \Im w \right] 
\left[\vartheta_1 (w - \bar w)\right]^{-q^2/(4\pi)} \, .
  \nn
\eea
We use 
\be
\vartheta_1' (w - \bar w) \left[
\vartheta_1 (w - \bar w)\right]^{-q^2/(4\pi) -1}
=  \frac{2\,i\, \pi}{q^2} \frac{\partial}{\partial \Im w} 
\left[\vartheta_1 (w - \bar w)\right]^{-q^2/(4\pi)} \, ,   
\ee
after which, the integrals over $\Im w$ give (at the leading order in
$q^2\rightarrow 0$)
\be
\int_0^\ell  \: \partial^2_{w} \ln \vartheta_1(w-\bar{w})
\, \di  \Im w = - \pi \, ,
\ee
\be
\int_0^\ell \Im w 
\frac{\partial}{\partial \Im w} 
\left[\vartheta_1 (w - \bar w)\right]^{-q^2/(4\pi)} \di \Im w  = 
-\ell
\ee
and
\be
\int_0^\ell 
\frac{\vartheta_1 ''(w - \bar w)}
{\left[\vartheta_1 (w - \bar w)\right]^{1 + q^2/(4\pi)}}\,  \di \Im w  = 
- \frac{8\,\pi^2}{q^2} \label{int3}
\ee
respectively. In eq.~(\ref{int3}), only the upper limit of integration 
contributes and we have used  the relation $\vartheta_1 (2 i \ell -y)
= e^{2\pi \ell + 2 \pi i y}\vartheta_1 (-y)$. 
 
At this point,
the integration over $\di^2 (z-w)$ 
in (\ref{f2f4}) gives the desired quadratic pole
 and we are left with the final result
\bea
a\, (r_\top ,b, v) & \simeq &  k_0^2\, \int 
 \frac{\di^{d\mbox{-}2} \vec q_\top}{(2\pi)^{d\mbox{-}2}}
\frac{1}{\vec q_\top^{\;2}}\:  e^{i \vec r\cdot \vec q_\top} \nn \\
& & \times \int_0^\infty \di \ell
\left\{ \frac{11}{4}\, v^4 - 2\, v_\top^2\, v^2 
-  b_z^2\,\frac{v^4}{4\ell} \right\} 
\ell^{-(d-1)/2} e^{-b^2/2\ell}\, , \label{last} 
\eea 
where we have dropped overall factors, the knowledge of which would
require fixing the absolute normalization of the string amplitude.
As discussed in the introduction, the first non-vanishing term is
of order $O(v^4)$ and its functional behavior in $b$
is independent from $\alpha'$.

Notice that, because of the three-body kinematics, the amplitude
(\ref{last}) is $O(v^4)$, whereas the interaction of two D-particles without
shock wave is $O(v^3)$~\cite{Bachas}.

A final comment.
The first term on the right-hand side of eq.~(\ref{ferm}) would also
give an
$\alpha'$-independent behavior in $b$, when summed over spin structure and
expanded at the order $O(v^2)$. However, since this term is 
proportional to $(\vec q_\top \cdot \vec v)^2/q_\top^2$, it is
sub-leading for $r_\top \rightarrow \infty$, and we do not keep it into account
in this paper.

\section{The matrix model in the shock wave}      

The action in the matrix model is given by the dimensionally reduced
$d=10$ and $N=1$ super Yang-Mills action. We are going to compute the one-loop
contribution to the effective action in a suitable background. 

The gauge-fixed bosonic action is
\be
S_{B} = \int \di t \: \left[ \: \frac{1}{2} \, \Tr F_{\mu\nu}F^{\mu\nu} + 
\Tr (D^B_\mu A^\mu )^2 
\right] \, , \label{sB} 
\ee
where 
\be
F_{0i} = \partial_t X_i 
\quad \mbox{and}
\quad F_{ij} = i\,[X_i, X_j] \, ,
\ee
since
\be
A^i \equiv X^i  \quad \mbox{and} \quad
A^0 =  0 \, .
\ee
The action (\ref{sB}) is expanded around the
classical  background field $B_\mu$ by separating
$A_\mu$ into $B_\mu$ and the fluctuation $X_\mu$ which is  
integrated out in the path integral.
The background covariant derivative is defined by
\be
D_\mu^B A_\nu = \partial_\mu A_\nu + i\,[B_\mu, A_\nu] \, .
\ee

The ghost action is
\be
S_C = \int \di t\: \Tr \bar{C} (D^B)^2 C \, ,
\ee
where the covariant derivative $D$ can be taken equal to $D^B$ 
because we are only interested in the quadratic part of the ghost
action.

Similarly, the
fermionic action is given by
\be
S_{F} = \int \di t\: \Tr \bar{\Psi} \Gamma^\mu D_\mu^B \Psi \, .
\ee

For the case of two D-particles, the fields
\be
X^i =  X^i_a \, \frac{\tau^a}{2} \, , \quad
C = C_a  \, \frac{\tau^a}{2} \quad \mbox{and} \quad
\Psi = \psi_a  \, \frac{\tau^a}{2} 
\ee
take value in the space of the gauge group $SU(2)$ (and $\tau^a$ are Pauli's
matrices).

In the eikonal approximation, and before introducing
the shock wave, the background field $B$ is taken to be 
\be
\vec B = \left( \begin{array}{cc} \vec v_1\, t + \vec b_1 & 0 \\
                                   0 &  \vec v_2\, t + \vec b_2
                 \end{array} \right) \quad \mbox{and} \quad  B_0=0 \, ,
\ee
where $\vec v_i$ and $\vec b_i$  
are the velocities and positions of the two D-particles.
In the frame of reference where 
$\vec v_1 + \vec v_2 = 0$ and $\vec b_1 + \vec b_2 =0$, we can then take
\be
\vec B = \left( \vec v\, t + \vec b \right) \frac{\tau_3}{2} \, ,
\ee
where now $\vec v_1 =- \vec v_2 = \vec v/2$,
$\vec b_1 =- \vec b_2 = \vec b/2$; $v$ and $b$ are the relative
velocity and distance in the motion of the two D-particles.
As we did for the string case, we have thus factorized out the motion of
the center of mass. Notice that the action of the matrix-valued background
field on the matrix valued quantum field $\phi$ (where $\phi$ can be 
$A^\mu$, $C$ or $\psi$) is $B \circ \phi = [B,\phi]$.

\subsection{Kinematics}

The space-time of the shock wave moving along $z$ from right to left
is described by the metric element
\be
\di s^2 = 2 \,\di U\, \di V + 	h \, \di U^2  + 
\di \vec x^{\,2}_{\top} \label{ds2}
\ee
where the light-cone variables $(U,V)$ are defined to be
\be
U \equiv (z +t)/\sqrt{2}\quad \mbox{and} \quad V  \equiv (z -t)/\sqrt{2} \, .
\ee
The coefficient $h$ is given by
\be
h = f(\vec r_\top) \delta(U) \, ,
\ee
where $\vec r_\top$ is
the distance from the center of mass of the two D-particles to the shock wave.
For a source graviton in $d=10$ and with an energy $k_0$
\be
f(\vec r_\top) =  \frac{8 \sqrt{2}}{\pi^3} 
\frac{G_N k_0}{r_\top^6}  \, , \label{f} 
\ee
where $G_N$ is Newton's constant.

From now on, we take $U$ as the evolution variable.

The metric components
must be thought as a matrix of $SU(2)$ depending on the
positions $\vec Y_i$ of the two D-particles. However,
\be
\left( \begin{array}{cc}  f(\vec Y_1) \delta(U) & 0 \\
                          0 &  f(\vec Y_2) \delta(U)
        \end{array} \right) =
\left( \begin{array}{cc}  f(\vec r_\top) \delta(U) & 0 \\
                          0 &  f(\vec r_\top) \delta(U)
        \end{array} \right) \left[1  + O\left(\frac{b_z}{r_\top}\right)
\right] \, , 
\ee          
and, for $b_z \ll r_\top$, we can neglect the higher order terms and take
the metric to be proportional to the identity matrix.

Because of the non-flat external metric, care must be taken in
lowering and rising indices.
In particular, in going from the background
gauge fields with lower indices to
the background coordinates with upper indices, we must use
\be
B_U = h\, B^U + B^V \quad \mbox{and} \quad B_V = B^U
\ee
while $\vec B_\top = \vec B^\top$.

We now fix $ B^U$, $B^V$ and $\vec B^\top$, by taking into account the 
trajectories of the D-particles. 

The trajectory of a particle moving in the shock-wave
background  is (see the appendix) 
\be
\left\{ \begin{array}{rcl} V & =& U\, w + w_0 - \frac{1}{2} f \theta (U) \\
                         \vec x_\top & =& \vec v_w\, U + \vec b_w \, .
        \end{array} \right.  \label{15}
\ee

By substituting $U=(z+t)/\sqrt{2}$ in (\ref{15}), we reproduce the 
D-particle trajectories (for $i=1,\,2$) before the shock 
 $z^{\,(i)}=v_{z}^{\,(i)}\, t+b_{z}^{\,(i)}$ and 
$\vec x_{\top}^{\,(i)} =\vec v_{\top}^{\,(i)} t+\vec b_{\top}^{\,(i)}$,  
by the assignment 
\be
w^{(i)} = \frac{v_{z}^{(i)} -1}{v_{z}^{(i)}+1} \, , 
\quad w_0^{(i)} = \frac{\sqrt{2}\,b_{z}^{(i)}}{v_{z}^{(i)}+1} \, , \label{wi}
\ee
and
\be
\vec v_{w}^{\,(i)} = 
\vec v_{\top}^{\,(i)}\frac{\sqrt{2}}{v_{z}^{(i)} +1}\, , \quad
\vec b_{w}^{\,(i)} = 
\vec b_{\top}^{\,(i)} - \vec v_{\top}^{\,(i)} \frac{b_{z}^{(i)}}{v_{z}^{(i)}+1}
\label{vi}
\, .
\ee

The matrix-valued  
$\vec B_\top =(\vec x_{\top}^{\,(1)}-\vec x_{\top}^{\,(2)})\, \tau_3/2$ is thus
found to be
\be
\vec B_\top =\left( \frac{\sqrt 2\, \vec v_\top}{1-v^2_z/4}U+ \vec b_\top
         +\frac{\vec v_\top\, b_z\, v_z}{4-v^2_z} \right) \, \frac{\tau_3}{2}
\label{Btop} \, .
\ee
The part of the background proportional to the identity is
irrelevant for our computation concerning the relative motion
of the D-particles.

Next, we have 
\be
\frac{B^U_{(i)}+B^V_{(i)}}{\sqrt{2}}=z_{(i)}=\frac{U+V_{(i)}}{\sqrt 2}
\label{constr1}
\ee
and, since $B_0=0$ implying $B_U=B_V$, we impose
\be
h\, B^U_{(i)}+B^V_{(i)}=B^U_{(i)}\, .
\label{constr2}
\ee
By solving these constraints at the leading order in $h$, 
we get the matrix-valued
\be
B_U=B_V=\frac{v_z\,U+b_z/\sqrt 2}{1-v^2_z/4} 
        \left( 1+\frac{h}{2} \right)\, \frac{\tau_3}{2} \, .
\label{BU}
\ee

Notice the absence of the shift in the trajectory, that is present in
the one particle problem; it cancels out in the two-body 
relative motion.
Also, we will see that we only need the leading first-order terms
in the velocity in (\ref{Btop}) and (\ref{BU}). 

Derivatives in the curved background are often complicated. 
However, an important property of the shock-wave metric is that, 
since $\partial_V g^{V\lambda} = \partial_V \delta(U) = 0$, we have that
\be
D_\nu^B g^{\nu\lambda} = g^{\nu\lambda} D_\nu^B 
\ee
with great simplifications in the computation.
Moreover, $\sqrt{g} =1$ and the covariant derivative
is the usual one. The derivative 
\be
\partial_U = - \partial_V \label{partials}
\ee
because the fields do not depend on $z$.

Similarly, for the fermions we can pass the covariant derivative $D_\mu^B$
through the $\Gamma$-matrices, and use 
\be
\Tr_{\!\!\gamma} \ln(\Gamma_\mu D^\mu) = \frac{1}{2}\:
\Tr_{\!\!\gamma} \ln(\Gamma_\mu D^\mu)^2=  \frac{1}{2}\: \Tr_{\!\!\gamma} \ln 
\left[ D^2 + \frac{1}{2} \Sigma_{\mu\nu}F^{\mu\nu} \right]
\label{traccia}
\ee
even in the shock-wave space-time (the spin connection is zero). 
Eq.~(\ref{traccia}) 
simplifies the evaluation of the
fermionic path integral.

\subsection{The quadratic action}

Because we are interested in a one-loop computation,
we need only the part of the action that is quadratic
in the fields. The bosonic action is then
\be
S_B =  \int \di U\: \Tr X^\mu \left[ -\delta^\nu_\mu D^2 - 2 i 
F^{\nu}_{\mu} \right] X_\nu \label{Sb} \, .
\ee

In eq.~(\ref{Sb}), the background field strength $F^\mu_\nu$
 is independent from $U$. As we shall see, because the result is
already proportional to $F^4$, we only need the part of $F$ that 
is leading order in $v$ and which is given by ($\mu,\nu = U, V, \top$)
\be
F_{\mu}^{\nu} = \left( \begin{array}{ccc} 2v_z & 0 & \sqrt{2} v_\top \\
                              0 & -2v_z & -\sqrt{2} v_\top  \\
                              \sqrt{2} v_\top & - \sqrt{2} v_\top & 0
              \end{array} 
\right) + h \, \left( \begin{array}{ccc} v_z & -2 v_z & 0 \\
                              0 & -v_z & 0 \\
                              0 & -\sqrt{2} v_\top & 0
              \end{array} 
\right) . \label{F}
\ee
Notice that in writing (\ref{F}), we used $\partial_V h =0$ as well as 
$\partial_U h \simeq 0$ (consistently with (\ref{partials}))
since terms proportional to $\delta'(U)$ would eventually multiply functions
of $U^2$ in the final amplitude and would not contribute.
 
In order to compute the amplitude for the
scattering of the two D-particles in the external field of the shock wave,
we need
\bea
a\, (r_\top ,b,v)  &=& \ln \int 
[\di X] [\di \bar C ] [ \di C] [\di \Psi] [\di \bar \Psi] e^{-S_B-S_C-S_F} =
 - \frac{1}{2} \Tr \ln \left(-\delta^{\nu}_{\mu} D^2 - 2 i F^{\nu}_{\mu}  \right) \nn \\ 
&& + \Tr \ln \left( -D^2 \right) 
+ \frac{1}{4} \Tr \ln \left(- D^2 - \frac{1}{2} \Sigma_{\mu\nu}F^{\mu\nu}
\right) \, .  \label{aMT}
\eea

The operator traces can be written in terms of the Schwinger
representation by introducing a parameter $s$. The amplitude (\ref{aMT}) thus
becomes
\be
a\, (r_\top ,b,v) = a_B + a_F \, ,
\ee
where
\be
a_B = - \frac{1}{2} \int \di U \:
\int_0^\infty \frac{\di s}{s}\: \lim_{U_{1,2} \rightarrow U} 
{\cal W}_g (s,U_1,U_2)\,
\left( \Tr_{\!\!\mu\nu}\, e^{2 i s F^\nu_\mu} - 2 \right) \label{aB}
\ee
for the bosonic part, and
\be
a_F = + \frac{1}{4} \int \di U \:
\int_0^\infty \frac{\di s}{s}\: \lim_{U_{1,2} \rightarrow U} 
{\cal W}_g (s,U_1,U_2)\,
\left( \Tr_{\!\!\gamma}\, e^{s \Sigma_{\mu\nu}F^{\mu\nu}/2} \right)
\label{aF}
\ee
for the fermionic one.

In (\ref{aB}) and  (\ref{aF}) we have separated between parenthesis
the part that does depend on $U$ and defined
\be
{\cal W}_g (s,U_1,U_2) = \langle U_1 |\:
e^{s D^2} | U_2 \rangle \, , \label{kernel}
\ee
the kernel of the scalar propagator in the shock wave space-time, where, to
leading order in $h$, $D^2$ is given by
\bea
D^2& =& 2 \,\partial_U\partial_V - 2 B_U B_V - \vec B_\top\cdot
\vec B_\top + h\, B_V B_V -
h\, \partial_V^2 \nn \\
&=& - 2\, \partial^2_U - 2 v^2 U^2 - b^2 +
h \left[ - \partial^2_U - \frac{b_z^2}{2}  \right] \, . \label{D2}
\eea

In writing (\ref{D2}), we have used (\ref{Btop}), (\ref{BU})  and (\ref{partials})
 after replacing for $V$ and $x_\top$
the values on the trajectory. In the the term proportional to $h$, we have
neglected contributions proportional to $v$ because this
part  is going to be multiply by $v^4$. The operator $D^2$ acts on the
components $\phi_a$ of the matrix-valued field $\phi=\phi^a\, \tau_a/2$
and $B_U$, $B_V$
and $B_\top$ in eq.~(\ref{D2}) are the coefficients 
in front of $\tau_3/2$ in eq.~(\ref{BU}) that remain
after performing the trace over the gauge group.
Terms linear in $B_\mu$ do not contribute to the
trace since $B \propto \tau_3$.

The kernel (\ref{kernel}) 
can be expanded around the flat space-time $(h=0)$ part:
\be
{\cal W}_g (s,U) \equiv \lim_{U_1, U_2 \rightarrow U} {\cal W}_g (s, U_1, U_2)
= {\cal W}_\eta (s,U) + h\, \Omega(s,U) + O(h^2) \, ,
\ee
where
\be
{\cal W}_\eta (s,U)= \lim_{U_{1,2}\rightarrow U} 
\langle U_1 | e^{- s (\partial_U^2 + 2 v^2 U^2 + b^2)} | U_2 \rangle
\ee
is just  the kernel for the
harmonic oscillator, that is
\be
{\cal W}_\eta (s,U) =  \sqrt{\frac{v}{2 \pi \sin 4 v s}} e^{-s b^2} e^{v U^2 
(\cos 4 v s - 1)/\sin 4 v s} \, . \label{weta}
\ee

For the flat space-time case, (\ref{aMT})  
reproduces the known result~\cite{Bachas,4,BBPT}
for the phase shift of two interacting
D-particles. In fact, traces of odd powers of $F^\mu_\nu$ vanish
and we have that
\be
\Tr_{\!\!\mu\nu}\, e^{2 i s F^\nu_\mu} -2 = 
(10-2) + 2 \left( \cos 4\, vs -1 \right) 
\ee
and
\be
 \Tr_{\!\!\gamma}\, e^{s \Sigma_{\mu\nu}F^{\mu\nu}/2} = 16 \cos 2\,vs \, ,
\ee
since $\Tr_{\!\!\mu\nu}\, F^2 = 8 v^2$ and
\be 
\Tr_{\!\!\gamma}\, \left(\Sigma_{\mu\nu}F^{\mu\nu}\right)^2 = 
- (16 \times 2) \Tr_{\!\!\mu\nu}\, F^2 \, , \label{tr2}
\ee
where the factor 16 comes from the 
Dirac trace. 

By taking the traces in (\ref{aB}) and  (\ref{aF}) we thus find
\be 
a\, (b,v) \simeq \int_0^\infty \frac{\di s}{s}\: \int \di U \:
{\cal W}_\eta (s,U)\,
\Bigr[ 4 \cos2\, v s - \cos 4\, vs - 3 \Bigr] \, , \label{flat0}
\ee
where
\be
\int \di U\: {\cal W}_\eta (s,U) = \frac{e^{-s b^2}}{2 \sin 2\,sv} \, .
\label{flat}
\ee

In the light-cone formalism, we reproduce the formulas of
refs.~\cite{Bachas,4,BBPT} with $2\,v$ in the place of $v$ there. 
The  different factor in front of $v$ is absorbed in the overall 
normalization since the use
of these formulas for string theory makes sense only up to $O(v^4)$
(actually, in the flat case, $O(v^3)$, due to eq.~(\ref{flat})). 

As we shall see, contrary to the flat space-time case, the amplitude 
in the shock-wave background is proportional to $h=f(r_\top)\delta(U)$,
and therefore the integration over $U$ yields 
$\int \di U\, h \, {\cal W} \simeq f(r_\top) e^{-sb^2}$ instead
of (\ref{flat}). In agreement with the string amplitude (\ref{last}),
the amplitude in the curved background
is thus $O(v^4)$ rather than $O(v^3)$, as in the flat case.

\subsection{The scattering amplitude}

In order to compute (\ref{aMT}) in the
non-flat metric, we must expand the exponential functions
in (\ref{aB}) and (\ref{aF}) in powers of $h$. To the leading order in $h$,
 we have that
\be
\Tr_{\!\!\mu\nu} F^2 = 8 \left( v^2 + h\, v_z^2 \right) \, , \quad
 \Tr_{\!\!\mu\nu} F^4 = 32 \left( v^4 + 2 h\, v_z^2 v^2 \right) \, .
\ee
As in the flat-space case above, (\ref{tr2}) holds together with
\be
\Tr_{\!\!\gamma}\, \left( \Sigma_{\mu\nu}F^{\mu\nu} \right)^4 = 
- (16\times 16) \Tr_{\!\!\mu\nu} F^4
\ee
and it is true in general that $\left(\Tr F^2 \right)^2 = 2 \Tr F^4$.

We can thus expand to fourth order 
the exponential functions and write that
\be
- \frac{1}{2}\: \left( \Tr_{\!\!\mu\nu}\, e^{2 i s F^\nu_\mu} - 2 \right)
+ \frac{1}{4}\: \left( \Tr_{\!\!\gamma}\, 
e^{s \Sigma_{\mu\nu}F^{\mu\nu}/2} \right)
 = - \frac{s^4}{4} \Tr F^4 \, , \label{traces}
\ee
where the constant terms as well as those quadratic in the velocity have
cancelled as it happens in the flat-metric case.

By means of (\ref{traces}), we can now
write (\ref{aMT}) as
\be
a\, (r_\top ,b,v) \simeq 
\int \di U\: \int \frac{\di s}{s} \:
\Bigl[ {\cal W}_\eta (s,U) + h\, \Omega(s,U) \Bigr]
 \left( - \frac{s^4}{4} \Tr F^4 \right) \, , 
\ee
where $\Omega(s,U)$ is given by
\bea
\Omega (s,U) &=& - \lim_{U_{1,2}\rightarrow U} s \left[ 
\partial_{U_1}^2 + \frac{b_z^2}{2}  \right] 
{\cal W}_\eta (s,U_1,U_2) \nn \\
& = &
-s \left[ \frac{1}{4s} +  \frac{b_z^2}{2} \right] {\cal W}_\eta (s,U) \, ,
\eea 
and ${\cal W}_\eta$ by (\ref{weta}).

Collecting all terms linear in $h$ yields, apart for an overall factor:
\be
a\, (r_\top ,b,v) \simeq 4\, \int \di U \:
f(\vec r_\top) \delta (U) \:\int \di s\: s^{5/2}  e^{-s b^2} 
\left\{  
\frac{7}{4}\, v^4 - 2\, v^2\, v_\top^2 -s\, b_z^2\, \frac{v^4}{2} \right\} \, ,
\label{fine}
\ee
where we have now taken (\ref{weta}) at $v=0$.

\subsection{Comparison with the string  result}

Going back to the string computation,
the amplitude (\ref{last}) is what we want to compare with 
the matrix theory result in (\ref{fine}).
After changing $\ell \rightarrow 1/2s$, setting $d=10$, and normalizing
the incoming and outgoing source-graviton states each
by the usual factor $1/\sqrt{k_0}$, we obtain
that the string amplitude is  (up to an overall constant)
\be
a\, (r_\top ,b, v) \simeq f(\vec r_\top)
\:  \int \di s\: s^{5/2}  e^{-s b^2} 
\left\{ \frac{11}{4}\, v^4 - 2\, v^2_\top\,v^2 
- s\,  b_z^2\,  \frac{ v^4}{2} \right\} \, . \label{ft}
\ee
since
\be
f(\vec r_\top)  \: \propto \: k_0\, \int 
 \frac{\di^{8} \vec q_\top}{(2\pi)^{8}}
\frac{1}{\vec q_\top^{\;2}}\:  e^{i \vec r_\top \cdot \vec q_\top} \, .
\ee

After integrating (\ref{fine}) over $U$,
we see that the matrix model result  (up to an overall constant)
becomes equal to that of string theory
 except for the numerical
 factor $7/4$ instead of $11/4$ in front of the $v^4$ term.
This missing term comes from the inclusion of the
Jacobian arising from the $\delta$-function constraining the D-particles
to lie on their respective trajectories, which in our
formalism amounts to implementing (\ref{constr1}) and
(\ref{constr2}) for both. 
In order to enforce this constraint,
 we replace the integration over $U$ by
\bea
\int \di U & \rightarrow & \int \di U \prod_{i=1,2} \int \di B^U_{(i)} 
\di B^V_ {(i)}
\, \delta\left( B^U_{(i)} + B^V_{(i)} - \sqrt{2}\, z^{(i)} \right)
\delta\left( B^U_{(i)} (1-h) - B^V_{(i)} \right)  \nn  \\
& & \simeq \frac{1}{4}\, \int \di U \: (1 + h) \: \\
&& \times \prod_{i=1,2} \int \di B^U_{(i)} \di B^V_{(i)} 
\, \delta\left( B^U_{(i)} - 
\frac{1}{\sqrt{2}}\left(1 + \frac{h}{2}\right) z^{(i)} \right)
\delta\left( B^V_{(i)}  -
\frac{1}{\sqrt{2}}\left(1 - \frac{h}{2}\right) z^{(i)}\right) \, .\nn
\eea
The extra (leading in $h$)
factor $1+h$ in front, after multiplication by $-s^4 \Tr F^4/4$,  
provides the missing term $4 s^4 v^4 h$  that adds a term $v^4$ to
(\ref{ft}) and
makes the matrix-model result agree with that of string theory, since
$7/4 + 1 = 11/4$, and we finally obtain that
\be
a\, (r_\top ,b,v) \simeq f(\vec r_\top) \: \frac{v^2}{b^7} \left[
\frac{11}{4}\,  v^2 - 2 \, v_\top^2 - \frac{7}{4}\,\frac{b_z^2}{b^2} \, v^2 
\right] \, . \label{ultima}
\ee
\acknowledgments

R.~I.\ would like to acknowledge the collaboration of C.\ N\'u\~{n}ez and
F.\ Hussain in the early stages of this work concerning the string computation,
and E.\ Gava, F.\ Morales and K.~S.\ Narain for
useful discussions. All three authors thank G.\ Ferretti for many discussions.

\appendix                             

\section{Notation and useful formulas}

We collect in this appendix formulas and results that we
used in the previous
sections but were too cumbersome to be inserted in the main text.

\subsection{String propagators}

The bosonic propagator is given by~(see, for instance, \cite{I1})
\be
\langle X^{\mu} (z)  X^{\nu} (w) \rangle = - \eta^{\mu\nu} G(z,w)
                - S^{\mu\nu} G(z,\bar{w}) \, ,
\ee
where the metric signature is given by
$\eta^{\mu\nu} = ( -1, \delta^{ij})$,
\be
G(z,w) = \frac{1}{4\pi} \ln \left| \frac{\vartheta_1 (z-w)}{\vartheta_1'(0)} \right|
    - \frac{\Im^2 (z-w)}{2 \Im \tau} \, ,
\ee
$S^{00} = -1$ and $S^{ij} = -\delta^{ij}$. Everywhere, we write
\be
\vartheta_i (x) \equiv \vartheta_i (x\,|\:\tau)
\ee
where $\tau = 2 i \ell$.

The fermionic propagator at velocity $v=0$ is
\be
\langle \psi^{\mu} (z)  \psi^{\nu} (w) \rangle_{S} = 
- \frac{\eta^{\mu\nu}}{4\pi}\: P_S
\ee
where 
\be
P_S =
\frac{\vartheta_{S} (z-w) \vartheta_1'(0)}{\vartheta_1 (z-w) \vartheta_S(0)}
\, .
\ee
At $v\neq0$ we have that (here the direction 1 is parallel and
the directions $i,j$ are orthogonal to $\vec v$)~\cite{I2}
\bea
\langle \psi^{0} (z)  \psi^{0} (w) \rangle_{S}& =&
  \langle \psi^{1} (z)  \psi^{1} (w) \rangle_{S} = \frac{1}{4\pi}\: Q_S\nn \\
\langle \psi^{0} (z)  \psi^{1} (w) \rangle_{S}& =&
  \langle \psi^{1} (z)  \psi^{0} (w) \rangle_{S} =  \frac{1}{4\pi}\: R_S \nn \\
\langle \psi^{i} (z)  \psi^{j} (w) \rangle_{S}& =&
\frac{\delta^{ij}}{4\pi}\: P_S \, ,
\eea
where the spin-structure-dependent functions are defined as
\bea
Q_S & =& \frac{1}{2} \left[ e^{v}\: 
\frac{\vartheta_S \left(w -iv/\pi\right) \vartheta_1'(0)}
{\vartheta_S \left(-iv/\pi\right) \vartheta_1(w)}
+ e^{-v}\: 
\frac{\vartheta_S \left(w +iv/\pi\right) \vartheta_1'(0)}
{\vartheta_S \left(-iv/\pi\right) \vartheta_1(w)} \right] \\
R_S & =& \frac{1}{2} \left[ e^{v}\:  
\frac{\vartheta_S \left(w -iv/\pi\right) \vartheta_1'(0)}
{\vartheta_S \left(-iv/\pi\right) \vartheta_1(w)}
- e^{-v}\:  
\frac{\vartheta_S \left(w +iv/\pi\right) \vartheta_1'(0)}
{\vartheta_S \left(-iv/\pi\right) \vartheta_1(w)} \right]  \, .
\eea

In performing the sum over the spin structure,
it useful to use Riemann's identity:
\be
\frac{1}{2}
\sum_S (\pm)_S\:
\vartheta_S (u) \vartheta_S (v)  \vartheta_S (w) \vartheta_S (s)  =  
\vartheta_1 (u_1) \vartheta_1 (v_1)  \vartheta_1 (w_1) \vartheta_1(s_1) \, ,
\ee
where
\bea
u_1 = (u+v+w+s)/2 & \quad & v_1= (u+v-w-s)/2 \nn \\
w_1= (u-v+w-s)/2 & \quad & s_1= (u-v-w+s)/2 \, .
\eea

\subsection{Shock-wave metric}

The space-time of the shock wave is defined by the
metric element (\ref{ds2}) which gives
a metric tensor (its signature is such that $\eta_{00} = -1$) 
with components:
\be
g_{UU} = h \, , \quad
g_{UV} = 1 \, , \quad
g_{VV} =  0 \, , \quad
g_{ij} = \delta_{ij} 
\ee
and
\be
g^{UU} = 0 \, ,\quad
g^{UV} = 1 \, ,\quad
g^{VV} =  -h \, ,\quad
g^{ij} = \delta^{ij} \, , 
\ee
where $h=f(r_\top)\delta(U)$.

Similarly, the einbein, necessary in writing the fermionic action, is
given by
\be
e^U_U = e^V_U = 1 \, ,\quad
e^U_V = 0 \, ,\quad
e_U^V = \frac{h}{2} \, ,\quad
e^j_i = \delta^i_j
\ee  
where $e^a_\mu \eta_{ab} e^b_\nu = g_{\mu\nu}$. The spin connection is
zero: $\omega_{ab\mu} = 0$.

The action for the motion of one particle in the shock-wave metric is
found by varying the metric line element (\ref{ds2}) and
it is given by
\be
S = \int \di s \left[ 2 \,\frac{\di U}{\di s} \frac{\di V}{\di s} 
+ f \delta(U)  \left( \frac{\di U}{\di s} \right)^2 + 
 \left( \frac{\di \vec x_\top}{\di s} \right)^2 \right] \, .
\ee
The trajectory is therefore
\be
\left\{ \begin{array}{rcl} V & =& U\, w + w_0 - \frac{1}{2} f \theta (U) \\
                          U & =& s \\
                         \vec x_\top & =& \vec v_w\, U + \vec b_w \, ,
        \end{array} \right. \label{a.15}
\ee
where the coefficients are given in section~3 by eqs.~(\ref{wi}) and
(\ref{vi}).
\clearpage
\renewcommand{\baselinestretch}{1}

\end{document}